\documentclass[runningheads]{llncs}
\usepackage[T1]{fontenc}
\usepackage{graphicx}
\newcommand{\tabincell}[2]{\begin{tabular}{@{}#1@{}}#2\end{tabular}}
\usepackage{caption}
\usepackage{stfloats}
\usepackage{multirow}
\usepackage{times}
\usepackage{epsfig}
\usepackage{amsmath}
\usepackage{amssymb}
\usepackage[ruled,linesnumbered]{algorithm2e}
\usepackage{graphicx}
\usepackage{algorithmic}
\usepackage{threeparttable}
\usepackage{subfigure}
\usepackage{ifpdf}

\usepackage{booktabs}
\usepackage{threeparttable}
\usepackage{bbding}
\usepackage{pifont}
\newcommand{\cmark}{\ding{51}}%
\newcommand{\xmark}{\ding{55}}%
\usepackage{color}

\usepackage{makecell} 

\begin{document}
\title{Time-Frequency Transformer: A Novel Time Frequency Joint Learning Method for Speech Emotion Recognition}

\titlerunning{Time-Frequency Transformer}
%
\author{Yong Wang$^{*,}$\inst{1} \and
Cheng Lu$^{*,}$\textsuperscript{\ding{41}}$^,$\inst{2,3} \and
Yuan Zong\textsuperscript{\ding{41}}$^,$\inst{2,3} \and
Hailun Lian \inst{1,3} \and
Yan Zhao \inst{1,3} \and
Sunan Li \inst{1,3}
}
%
\authorrunning{Y. Wang et al.}
%
\institute{School of Information Science and Engineering, Southeast University, Nanjing 210096, China \and
Key Laboratory of Child Development and Learning Science of Ministry of Education, Southeast University, Nanjing 210096, China \and
School of Biological Science and Medical Engineering, Southeast University, Nanjing 210096, China \\
\email{\{cheng.lu, xhzongyuan\}@seu.edu.cn}. \\ $^*$ Equal Contributions. \textsuperscript{\ding{41}} Corresponding Authors.}

%
\maketitle              
\begin{abstract}
In this paper, we propose a novel time-frequency joint learning method for speech emotion recognition, called Time-Frequency Transformer. Its advantage is that the Time-Frequency Transformer can excavate global emotion patterns in the time-frequency domain of speech signal while modeling the local emotional correlations in the time domain and frequency domain respectively. For the purpose, we first design a Time Transformer and Frequency Transformer to capture the local emotion patterns between frames and inside frequency bands respectively, so as to ensure the integrity of the emotion information modeling in both time and frequency domains. Then, a Time-Frequency Transformer is proposed to mine the time-frequency emotional correlations through the local time-domain and frequency-domain emotion features for learning more discriminative global speech emotion representation. The whole process is a time-frequency joint learning process implemented by a series of Transformer models. Experiments on IEMOCAP and CASIA databases indicate that our proposed method outdoes the state-of-the-art methods.

\end{abstract}
\section{Introduction}

Speech Emotion Recognition (SER) aims to use computers to automatically analyze and recognize the emotional state in human speech\cite{schuller2013computational}, which has become a hotspot in many fields, e.\,g., affective computing and Human-Computer Interaction (HCI) \cite{schuller2018speech}. For SER task, how to obtain a speech emotion representation with high discrimination and strong generalization is a key step to realize the superior performance of speech emotion classification \cite{akccay2020speech}, \cite{lu2022domain}.

In order to recognition speech emotions well, early SER works mainly combined some low-level descriptor (LLD) features or their combinations \cite{stuhlsatz2011deep}, e.\,g., Mel-Frequency Cepstral Coefficients (MFCC), Zero-Crossing Rate, and Pitch, with classifiers \cite{schuller2009acoustic}, e.\,g., k-Nearest Neighbor (KNN) and Support Vector Machine (SVM), for emotion prediction. Furthermore, with the rapid development of deep learning, high-dimensional speech emotion features generated by deep neural networks (DNN), e.\,g., Convolutional Neural Network (CNN) \cite{abbaschian2021deep} and Recurrent Neural Network (RNN) \cite{wang2020speech}, have emerged on SER and achieved superior performance. Currently, the input features of DNNs are mainly based on spectrogram, e.\,g., magnitude spectrogram \cite{mao2014learning} and Mel-spectrogram \cite{lu2022speech}, which are the time-frequency representations of speech signals.

The time and frequency domains of the spectrogram contain rich emotional information. To excavate them, a practical approach is joint time-frequency modeling strategy with the input features of spectrograms \cite{lu2022speech}, \cite{akccay2020speech}. Among these methods, combining CNN and RNN structure, i.\,e., CNN$+$LSTM, is a classic method, which utilizes CNN and RNN to encode the information in frequency and time domains, respectively. For instance, Satt et al. \cite{satt2017efficient} combined CNN with a special RNN (i.\,e., LSTM) to model the time-frequency domain of emotional speech. Wu et al. \cite{wu2019speech} proposed a recurrent capsules network to extract time-frequency emotional information from the spectrogram.

Although current time-frequency joint learning methods have achieved certain success on SER, they still suffer from two issues. The first one is that they usually shared the modeling both in time and frequency domains, ignoring the specificality of the respective domains. For instance, the time-frequency domain shares a uniform size convolution kernel in CNN \cite{mao2014learning} and a uniform-scale feature map is performed on the time-frequency domain in RNN \cite{satt2017efficient}. Therefore, separate modeling of time-domain and frequency-domain information should be considered to ensure the specificity and integrity of the encoding of time-frequency domain information. The other issue is that only some low-level feature fusion operations (e.\,g., splicing and weighting) are adopted in time frequency joint learning process, leading to poor discriminativeness of fusion features \cite{lu2022speech}. This indicates that the effective fusion of emotional information in the time-frequency domain is also the key for time-frequency joint learning.

\begin{figure}
  \centering
  \includegraphics[width=3.2in,height=3.2in]{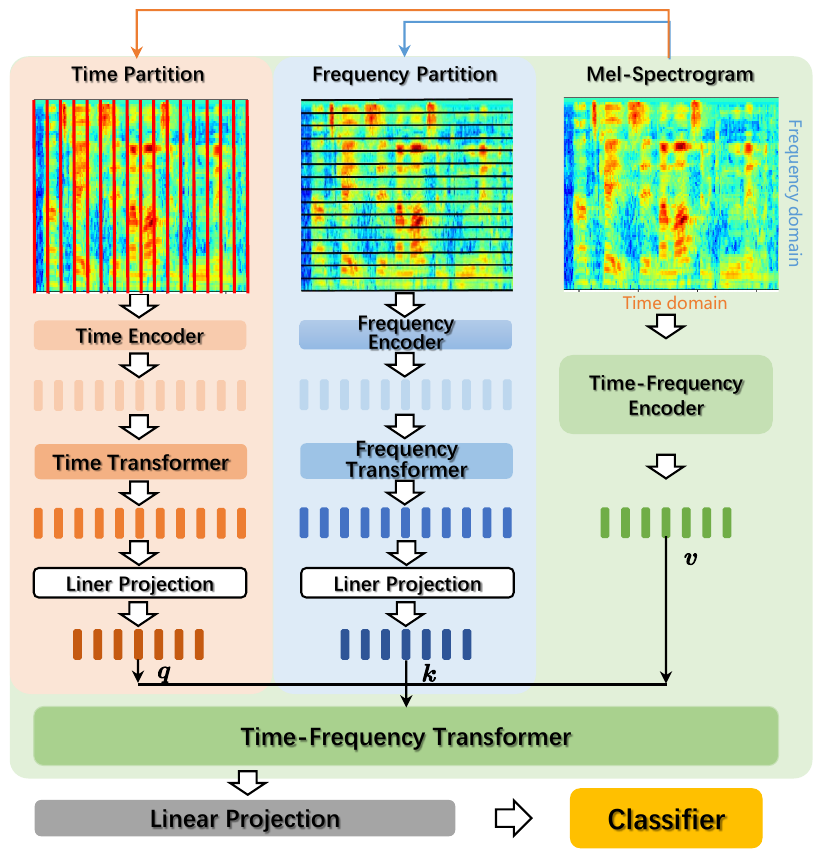}

  \caption{Overview of Time-Frequency Transformer for speech emotion recognition. It mainly consists of three modules, i.\,e., Time Transformer module (light orange background), Frequency Transformer module (light blue background), and Time-Frequency Transformer module (light green background).}
  \label{fig:framework}
\end{figure}

To cope with the above issues, we propose a novel Transformer-based time frequency domain joint learning method for SER, called Time-Frequency Transformer, which consisting of three modules, i.\,e., Time Transformer module, Frequency Transformer module, and Time-Frequency Transformer module, as shown in Figure. \ref{fig:framework}. Firstly, Time Transformer module and Frequency Transformer module are designed to model the local emotion correlations between frames and inside frequency bands respectively, which can ensure the integrity of the emotion information modeling in both time and frequency domains. Then, we also propose a Time-Frequency Transformer module to excavate the time-frequency emotional correlations through the local time-domain and frequency-domain emotion features for learning more discriminative global speech emotion representation. The whole process is a time-frequency joint learning process. Overall, our contributions can be summarized as the following three points:
\begin{itemize}
	\item We propose a novel time frequency joint learning method based on Transformers (i.\,e., Time-Frequency Transformer), which can effectively excavate the local emotion information both in time frames and frequency bands to aggregate global speech emotion representations.
	\item We propose a Time Transformer and Frequency Transformer to ensure the integrity of modeling time-frequency local emotion representations.
	\item Our proposed Time-Frequency Transformer outperform on the state-of-the art methods on IEMOCAP database and CASIA database.
\end{itemize}

\section{Proposed Method}
In this section, we will introduce our proposed Time-Frequency Transformer shown in Figure. \ref{fig:framework}, including three modules, i.e., Time Transformer module, Frequency Transformer module and Time-Frequency Transformer module.

\subsection{Time Transformer Module}
The role of Time Transformer is to capture the local emotion correlations across time frames for time domain encoding of emotional speech. 
This module utilizes the time encoder to reduce the dimensionality of the input feature $\boldsymbol{x}$ of the model and extract part of the emotional information in time-domain of $\boldsymbol{x}$. Then, MSA mechanism of the Transformer encoder is used to calculate the emotional correlation between frames, making the model focus on emotion related frames and reducing the impact of emotion unrelated frames on speech emotion recognition.

\indent Basically, the time encoder consists of a 2D convolutional layer, a 2D batch normalization layer, an activation function layer, a 2D convolutional layer, a 2D batch normalization layer and an activation function layer in sequence. We convolute the input spectrogram feature $\boldsymbol{x}\in\mathbb{R}^{b \times c \times f\times d}$ frame by frame and extract temporal emotional information. Each convolution layer is followed by a batch normalization layer and an activation function layer to speed up model training and improve the model's nonlinear representation capabilities. The output of the second activation function layer is $\boldsymbol{\hat{x}}^{\prime}\in\mathbb{R}^{b \times c1 \times (f/4)\times d}$, where \emph{b} and \emph{f} represent the number of samples selected for each training and the number of Mel-filter banks, \emph{c} and \emph{c1} represent the number of input channels of the spectrogram feature and the number of output channels of the last convolutional layer and d is the number of frames of the spectrogram feature.The process can be represented as
\begin{equation}
\boldsymbol{\hat{x}}^{\prime} = Act(BN(C_t(Act(BN(C_t(\boldsymbol{x}))))))
\end{equation}
where $C_t(\cdot)$ is the convolutional operation in each frame by a 2D convolutional layer, $BN(\cdot)$ and $Act(\cdot)$ is batch normalization and activation function operations respectively.

\indent We utilizes a Transformer encoder to perform temporal attention focusing on $\boldsymbol{\hat{x}}^{\prime}$. The Transformer encoder consists of a MSA sub-layer and a feed-forward neural network using residual connections.The Transformer encoder applies multiple attention heads to achieve the model parallel training. By dividing the input into multiple feature subspaces and applying a self-attention mechanism in the subspaces, the model can be trained in parallel while capturing emotional information.

\indent We take the mean value of the convolution output channel dimension of $\boldsymbol{\hat{x}}^{\prime}$ to get $\boldsymbol{s}^{\prime}\in\mathbb{R}^{b \times (f/4)\times d}$ and then transpose $\boldsymbol{s}^{\prime}$ to get $\boldsymbol{\hat{s}}^{\prime}\in\mathbb{R}^{b \times d\times (f/4)}$. The feature $\boldsymbol{\hat{s}}^{\prime}\in\mathbb{R}^{b \times d\times (f/4)}$ is used as the input of the time-domain Transformer encoder. We can represent the process as
\begin{equation}
\boldsymbol{m} =  LN(MSA(\boldsymbol{\hat{s}}^{\prime} )+\boldsymbol{\hat{s}}^{\prime})
\end{equation}
\begin{equation}
\boldsymbol{\hat{q}} =  LN(MLP(\boldsymbol{m})+\boldsymbol{m})
\end{equation}
where $MSA(\cdot)$, $LN(\cdot)$ and $MLP(\cdot)$ is MSA, layer normalization and feed-forward neural network respectively in a Transformer encoder. Besides, $\boldsymbol{m}$ is the output of $\boldsymbol{\hat{s}}^{\prime}$ after the MSA and layer normalization of a Transformer encoder and $\boldsymbol{\hat{q}}\in\mathbb{R}^{b \times d\times (f/4)}$ is the output of a Transformer encoder. The output feature $\boldsymbol{\hat{q}}$ will be used as one of the inputs of time-frequency Transformer module after linear mapping.
\subsection{Frequency Transformer Module}
This module has a similar structure to the time Transformer module, both containing a Transformer encoder. The difference between the two modules is that this module performs 2D convolutional operation on the input features $\boldsymbol{x}$ frequency band by frequency band, which is used for reducing dimension of feature and extracting frequency-domain emotional information. We use 2D convolutional operation $C_f(\cdot)$ on each frequency band of feature  $\boldsymbol{x}$. Then, we normalize the feature and activate the feature using the activation function. Finally get the output $\boldsymbol{\hat{x}}^{\prime\prime}\in\mathbb{R}^{b \times c1 \times f\times (d/4)}$ of the last activation function layer. The operations can be represented as
\begin{equation}
\boldsymbol{\hat{x}}^{\prime\prime} =  Act(BN(C_f(Act(BN(C_f(\boldsymbol{x}))))))
\end{equation}

\indent We take the mean value of the convolution output channel dimension of $\boldsymbol{\hat{x}}^{\prime\prime}$ to get $\boldsymbol{\hat{s}}^{\prime\prime}\in\mathbb{R}^{b \times f\times (d/4)}$. Then, we use a Transformer encoder to calculate the emotional correlation between frequency bands for $\boldsymbol{\hat{s}}^{\prime\prime}$, so that the model focuses on the emotional related parts in frequency domain, reducing the impact of emotional unrelated frequency bands on speech emotion recognition. The operations can be represented as
\begin{equation}
\boldsymbol{n} =  LN(MSA(\boldsymbol{\hat{s}}^{\prime\prime} )+\boldsymbol{\hat{s}}^{\prime\prime})
\end{equation}
\begin{equation}
\boldsymbol{\hat{k}} =  LN(MLP(\boldsymbol{n})+\boldsymbol{n})
\end{equation}
where $\boldsymbol{n}$ is the output of $\boldsymbol{\hat{s}}^{\prime\prime}$ after the MSA and layer normalization of the Transformer encoder and $\boldsymbol{\hat{k}}\in\mathbb{R}^{b \times f\times (d/4)}$ is the output of the Transformer encoder. After linear mapping, $\boldsymbol{\hat{k}}$ will be used as one of the inputs to the time-frequency Transformer module.
\subsection{Time-Frequency Transformer Module}
Time-Frequency Transformer Module aims to aggregate the local emotion encoding in time and frequency domains generated by Time Transformer and Frequency Transformer into the global emotion representation in time-frequency domain.
Therefore, the main function of this module is to use the time-frequency domain features of the speech that have been weighted by the MSA mechanism to use the Multi-head Attention(MA) mechanism \cite{chaudhari2021attentive} to further weight, so that the model pays more attention to the emotion-related segments in time-frequency domain, thereby improving accuracy of speech emotion recognition.

\indent First, We utilize the 2D convolution operation $C(\cdot)$ to encode $\boldsymbol{x}$ in the time-frequency domain. After that, we normalize and activate the feature to obtain $\boldsymbol{\hat{x}}\in\mathbb{R}^{b \times c1 \times (f/4)\times (d/4)}$ . The process can be represented as
\begin{equation}
\boldsymbol{\hat{x}} = Act(BN(C(Act(BN(C(\boldsymbol{x}))))))
\end{equation}

\indent Then, we take the mean value of $\boldsymbol{\hat{x}}$ in the convolution output channel dimension to get $\boldsymbol{v}\in\mathbb{R}^{b \times (f/4)\times (d/4)}$. We linearly map $\boldsymbol{\hat{q}}$ and $\boldsymbol{\hat{k}}$ to get $\boldsymbol{q}\in\mathbb{R}^{b \times (f/4)\times (d/4)}$  and $\boldsymbol{k}\in\mathbb{R}^{b \times (f/4)\times (d/4)}$ respectively.  The $\boldsymbol{q}$, $\boldsymbol{k}$ and $\boldsymbol{v}$ are used as the input of the CO-Transformer encoder. Compared with an original Transformer encoder, we replace the MSA sub-layer in the encoder with a MA sub-layer to obtain a CO-Transformer encoder. A CO-Transformer encoder is composed of a MA sub-layer and a feed-forward neural network using residual connection.The main difference between MSA and MA is that when doing attention calculations, the inputs $\boldsymbol{Q}$ (Query Vector), $\boldsymbol{K}$ (Keyword Vector), and $\boldsymbol{V}$ (Value Vector) of MSA are the same, but the inputs $\boldsymbol{Q}$ ,$\boldsymbol{K}$, and $\boldsymbol{V}$ of MA are different. The process can be represented as
\begin{equation}
\boldsymbol{p} =  LN(MA(\boldsymbol{q},\boldsymbol{k},\boldsymbol{v} )+\boldsymbol{v})
\end{equation}
\begin{equation}
\boldsymbol{y} =  LN(MLP(\boldsymbol{p})+\boldsymbol{p})
\end{equation}
where $MA(\cdot)$ is MA and $\boldsymbol{p}$ is the output of the input features after the MA and layer normalization of the CO-Transformer encoder. Besides, $\boldsymbol{y}\in\mathbb{R}^{b \times (f/4)\times (d/4)}$ is the output of the CO-Transformer encoder. The obtained $\boldsymbol{y}$ is the input of the classifier and finally obtain the predicted emotional category.

\indent The classifier consists of a pooling layer and a fully connected layer. The main function of the pooling layer is to reduce the feature dimension. The pooling layer takes the mean and standard deviation in the frequency-domain dimension of $\boldsymbol{y}$  and concatenates them to get $\boldsymbol{\hat{y}}\in\mathbb{R}^{b \times (d/2)}$, which can be represented as
\begin{equation}
\boldsymbol{\hat{y}} =  Pool(\boldsymbol{y})
\end{equation}
where $Pool(\cdot)$ is the pooling operation.

\indent We calculate the prediction probability of all emotions through a fully connected layer and ultimately obtain $\boldsymbol{\hat{y}}^{\prime}\in\mathbb{R}^{b \times c}$, where \emph{c} is the number of emotion categories of the corpus. We take the emotion with the largest prediction probability as the predicted emotion of the model and optimize the model by reducing the cross-entropy loss \emph{Loss} between the predicted emotion label $\boldsymbol{\hat{y}}^{\prime\prime}\in\mathbb{R}^{b \times c}$ and the true emotion label $\boldsymbol{z}\in\mathbb{R}^{b \times c}$. The operations can be represented as
\begin{equation}
\boldsymbol{\hat{y}}^{\prime} =  FC(\boldsymbol{\hat{y}})
\end{equation}
\begin{equation}
\boldsymbol{\hat{y}}^{\prime\prime} =  Softmax(\boldsymbol{\hat{y}}^{\prime})
\end{equation}
\begin{equation}
Loss = CrossEntropyLoss(\boldsymbol{\hat{y}}^{\prime\prime},\boldsymbol{z}))
\end{equation}
where $FC(\cdot)$, $Softmax(\cdot)$ and $CrossEntropyLoss(\cdot)$ is the fully connected layer operation, Softmax function \cite{dubey2022activation}  and Cross-Entropy loss function \cite{zhang2018generalized} respectively.
\section{Experiments}
\subsection{Experimental Databases}
Extensive experiments are conducted on two well-known speech emotion databases, \emph{i.e.}, $\textbf{IEMOCAP}$ \cite{busso2008iemocap} and $\textbf{CASIA}$ \cite{zhang2008design}.
\begin{itemize}
  \item $\textbf{IEMOCAP}$ is an audio-visual database released by the Sail Laboratory of the University of Southern California. This database consists of five dyadic sessions, and each session is performed by a male actor and a female actor in improvised and scripted  scenarios to obtain various emotions (angry, happy, sad, neutral, frustrated, excited, fearful, surprised, disgusted, and others). We select the audio samples in improvised scenario, including 2280 sentences belonging to four emotions (angry, happy, sad, neutral) for experiments.
  \item $\textbf{CASIA}$ is a mandarin database collected by the Institute of Automation of the Chinese Academy of Sciences. Four speakers are required to perform six different emotions, \emph{e.g.}, happiness, anger, fear, sadness, neutrality and surprise. A total of 1200 sentences are utilized in this experiment. The sampling rate of both databases is 16 kHz. The details of the above two databases are reported in Table~\ref{tab:description-of-databases}.
\end{itemize}



\begin{table}
\caption{Experimental Database Description}
\label{tab:description-of-databases}
\centering
\renewcommand{\arraystretch}{1.2}
\begin{tabular}{c|c|c|c}
\hline
Database  &  Language  &  Samples of Each Emotion                      & Total Samples \\
\hline \hline
IEMOCAP   &  English   & \tabincell{c}{\emph{angry} (289) \emph{happy} (284)\\ \emph{neutral} (1099) \emph{sad} (608) }   &  2280      \\
\hline
CASIA& Mandarin   &\tabincell{c}{\emph{angry} (200) \emph{fear} (200)\\ \emph{happy} (200) \emph{neutral} (200)\\ \emph{sad} (200) \emph{surprise} (200)}
& 1200       \\
\hline
\end{tabular}
\end{table}

\subsection{Experimental Protocol}
In the experiment, we follow the same protocol of the previous research \cite{lu2022speech} and adopt the Leave-One-Speaker-Out (LOSO) cross-validation for evaluation.

Specifically, for CASIA, when one speaker's samples are served as the testing data, the remaining three speakers' samples are used for training. Similarly, for IEMOCAP, we use one speaker's samples as the testing data and other speakers' samples as the training data. Moreover, since IEMOCAP contains 5 sessions, the leave-one-session-out cross-validation protocol (one session's samples as the testing data and four sessions' samples as the training data) is also a common way for evaluation \cite{bhosale2020deep}. Therefore, in Table 4, we also choose some methods using this protocol to compare with our proposed method.

In this paper, we choose the weighted average recall (WAR) \cite{schuller2009acoustic} and the unweighted average recall (UAR) \cite{stuhlsatz2011deep}, which are widely-used SER evaluation indicators,  to effectively measure the performance of the proposed method.

\subsection{Experimental Setting}
Before the feature extraction, we preprocess the audio samples by dividing them into small segments of 80 frames (20ms per frame). With this operation, samples are not only augmented, but also maintain the integrity of speech emotions. After that, we pre-emphasize the speech segments and the pre-emphasis coefficient is 0.97. Then, we use a 20ms Hamming window with a frame shift of 10ms to extract log-Mel-spectrogram, where the number of points of Fast Fourier Transform (FFT) and bands of Mel-filter are 512 and 80 respectively. Finally, the model input features of \emph{b}=64, \emph{c}=1, \emph{f}=80, \emph{d}=80 are obtained.

\indent Besides, the parameters of $C_t(\cdot)$, $C_f(\cdot)$, $C(\cdot)$ are shown in Table~\ref{tab:parameters-of-CNN}. The $BN(\cdot)$ and $Act(\cdot)$ denote the BatchNorm function \cite{ioffe2015batch} and the ReLU function \cite{dubey2022activation}, respectively. The parameters of the Transformer encoder used in our model are shown in Table~\ref{tab:parameters-of-Transformer}. The proposed method is implemented by Pytorch \cite{paszke2019pytorch} with NVIDIA A10 Tensor Core GPUs, which is trained from scratch with 1000 epochs and optimized by Adam optimizer \cite{adam2014method} with the initialized learning rate of 0.001.
\begin{table}
\caption{CNN Parameters Settings }
\setlength{\tabcolsep}{6mm}
\label{tab:parameters-of-CNN}
\renewcommand{\arraystretch}{1.2}
\centering
\begin{tabular}{c|c|c|c|c}
\hline
Operation        &  Out Channels      & Kernel Size              & Stride           & Padding  \\
\hline \hline
$C_t(\cdot)$     &  64                & (5,1)                    & (2,1)            & (2,0)      \\
\hline
$C_f(\cdot)$     &  64                & (1,5)                    & (1,2)            & (0,2)     \\
\hline
$C(\cdot)$       &  64                & (5,5)                    & (2,2)            & (2,2)     \\
\hline
\end{tabular}
\end{table}

\begin{table}
\caption{Transformer Encoder Parameters Settings}
\label{tab:parameters-of-Transformer}
\renewcommand{\arraystretch}{1.2}
\centering
\begin{tabular}{c|c|c|c}
\hline
Module        &  Embed Dimension      & Feed-forward Dimension               &  Attention Heads            \\
\hline \hline
Time Transformer     &  20                & 512                    & 2                  \\
\hline
Frequency Transformer     &  20                & 512                    & 2                 \\
\hline
Time-Frequency Transformer      &  20                & 1024                    & 4                 \\
\hline
\end{tabular}
\end{table}

\begin{table}[t]
\caption{Experimental Results on IEMOCAP and CASIA}
\label{tab:experimental-results}
\centering
\begin{tabular}{|c|c|c|c|c|}
\hline
\multirow{2}{*}{Database} & \multirow{2}{*}{Experimental Protocol} & \multirow{2}{*}{Comparison Method} & \multicolumn{2}{c|}{Accuracy(\%)} \\ \cline{4-5}
                          &                                            &                                                   & WAR  & UAR     \\ \hline \hline
\multirow{7}{*}{IEMOCAP}  & \multirow{7}{*}{\begin{tabular}[c]{@{}c@{}}Leave One Session/Speaker Out (LOSO) \\ (5 Sessions or 10 Speakers)\\\end{tabular}}
                                                                       & MSA-AL \cite{bhosale2020deep}       & 72.34           & 58.31   \\ \cline{3-5}
                          &                                            & CNN-LSTM \cite{satt2017efficient}    & 68.80           & 59.40   \\ \cline{3-5}
                          &                                            & STC-Attention \cite{guo2021representation}             & 61.32           & 60.43   \\ \cline{3-5}
                          &                                            & DNN-SALi \cite{mao2019revisiting}    & 62.28          & 58.02   \\ \cline{3-5}
                          &                                            & GRU-CNN-SeqCap \cite{wu2019speech}               & 72.73           & 59.71   \\ \cline{3-5}
                          &                                            & DNN-BN \cite{kim2019dnn}   & 59.7           & 61.4   \\ \cline{3-5}
                          &                                            & Ours                                      & \textbf{74.43}  & \textbf{62.90} \\ \hline

\multirow{6}{*}{CASIA}    & \multirow{6}{*}{\begin{tabular}[c]{@{}c@{}}Leave One Speaker Out (LOSO) \\ (4 Speakers)\end{tabular}}
                                                                       & GA-BEL \cite{liu2018speech}                    & 39.50           & 39.50     \\ \cline{3-5}
                          &                                            & ELM-DNN \cite{han2014speech}                  & 41.17           & 41.17     \\ \cline{3-5}
                          &                                            & LoHu \cite{sun2015weighted}                & 43.50           & 43.50     \\ \cline{3-5}
                          &                                            & DCNN-DTPM \cite{zhang2017speech}                 & 45.42           & 45.42     \\ \cline{3-5}
                          &                                            & ATFNN \cite{lu2022speech}                & 48.75           & 48.75     \\ \cline{3-5}
                          &                                            & Ours                                      & \textbf{53.17}  & \textbf{53.17} \\ \hline
\end{tabular}
\end{table}

\begin{figure*}[h]
\centering
\subfigure[Confusion matrix on IEMOCAP]{
\label{iemocap-cm}
\includegraphics[width=0.49\linewidth]{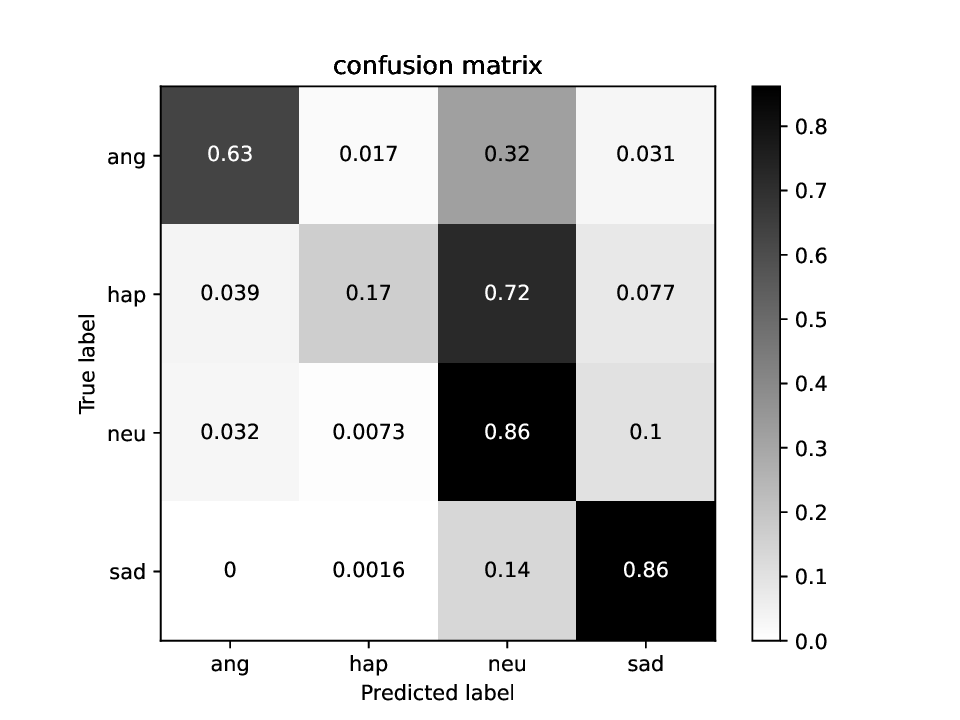}}
\subfigure[Confusion matrix on CASIA]{
\label{casia-cm}
\includegraphics[width=0.49\linewidth]{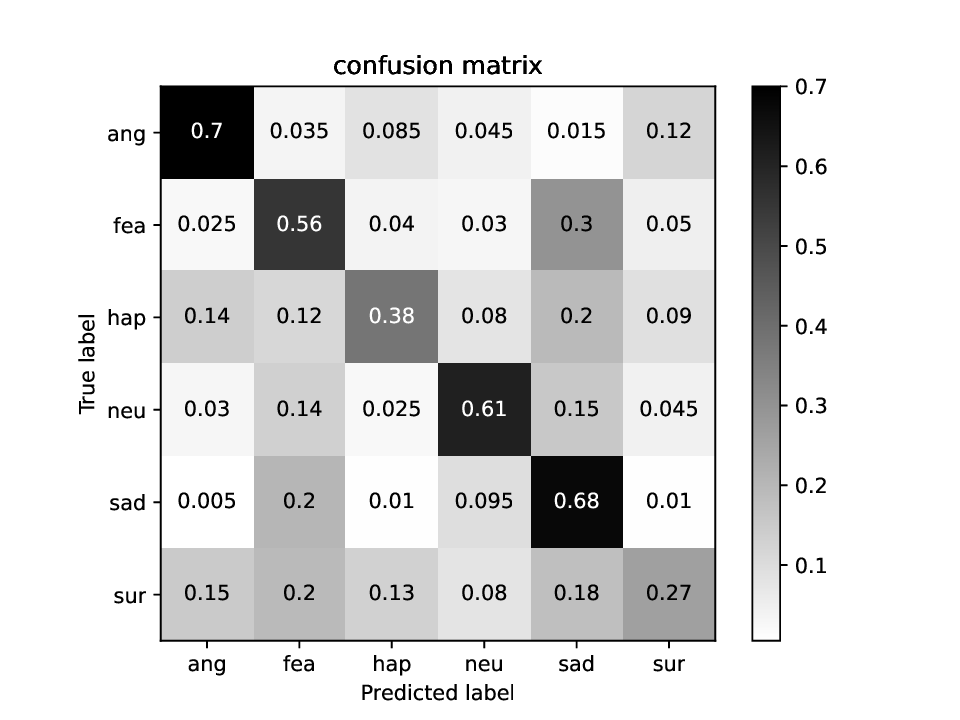}}

\caption{Confusion matrices on IEMOCAP and CASIA. }
\label{confusion-matrices}
\end{figure*}

\subsection{Experimental Results and Analysis}
\subsubsection{Results on IEMOCAP}
We selected some state-of-the-art methods for performance comparison with the proposed method, \emph{i.e.}, a model based on MSA that fuses acoustic and linguistic features (MSA-AL) \cite{bhosale2020deep} , a model that combines CNN with Long Short Term Memory (LSTM) and uses spectrogram as the input features (CNN-LSTM) \cite{satt2017efficient}, spectro-temporal and CNN with attention model(STC-Attention) \cite{guo2021representation}, a Deep Neural Network (DNN) method combining Gaussian Mixture Models (GMM) and Hidden Markov Models (HMM) using subspace alignment strategy (DNN-SAli) \cite{mao2019revisiting}, a method of using Gated Recurrent Unit (GRU) in CNN layers and combining with sequential Capsules (GRU-CNN-SeqCap) \cite{wu2019speech}, and a DNN method with Bottleneck features (DNN-BN) \cite{kim2019dnn}.

The experimental results are shown in Table~\ref{tab:experimental-results}. From the results, we can find something interesting. Firstly, our proposed Time-Frequency Transformer achieves the best performance on both WAR (74.43\%) and UAR (62.90\%) compared to other mentioned methods. Moreover, compared to all methods using Leave-One-Speaker-Out protocol, our proposed method achieves a promising increase over 10\% in term of WAR and 1.5\% in UAR.

\indent The confusion matrix of IEMOCAP is shown in Figure. \ref{iemocap-cm}. What we can observe first is that the proposed method exhibits a excellent performance in classifying specific emotions, \emph{e.g.}, \emph{angry}, \emph{neutral} and \emph{sad}. However, it is difficult for the proposed model to correctly recognize the emotion \emph{happy}. As shown in the figure, 72\% \emph{happy} samples are misclassified as \emph{neutral} and only 17\% are correctly classified. Obviously, it cannot be caused by the reason that \emph{happy} is more close to \emph{neural} than other emotions (negative emotions: \emph{anger} and \emph{sad}) since the possibility of \emph{neutral} samples being misclassified as \emph{happy} is only 0.73\%. This situation lead us to consider the other reason which is the unbalanced sample size. Since the number of \emph{happy} samples in IEMOCAP is only 284 which is the smallest among all emotions, the model cannot learn the unique emotional characteristics of \emph{happy} well. It may lead to this situation that the \emph{happy} samples are more likely to be mistaken for \emph{neutral}.

\subsubsection{Results on CASIA}
Some state-of-the-art methods that also use the LOSO protocol are used for comparison with the proposed method, including Genetic Algorithm (GA) combined with Brain Emotional Learning (BEL) model (GA-BEL) \cite{liu2018speech}, Extreme Learning Machine (ELM) combined with DNN (ELM-DNN) \cite{han2014speech}, weighted spectral features based on Local Hu moments (LoHu) \cite{sun2015weighted}, Deep CNN (DCNN) combined with a Discriminant Temporal Pyramid Matching (DTPM) strategy (DCNN-DTPM) \cite{zhang2017speech} and an Attentive Time-Frequency Neural Network (ATFNN) \cite{lu2022speech}.

The results on the CASIA database is shown in Table~\ref{tab:experimental-results}. It is obvious that our method achieves state-of-the-art performance among all algorithms. Specifically, our method obtains the best result on WAR (53.17\%) and UAR (53.17\%) than all comparison methods. Since the sample numbers of the 6 emotions of CASIA used in the experiment are balanced, WAR and UAR are equal. Besides, our results are not only the best, but also far superior to other methods. Even compared to ATFNN which is the second best method, the proposed method still obtain a over 4\% performance increase.

From the confusion matrix of CASIA in Figure. \ref{casia-cm}, it is obvious that the proposed method has a high recognition rate in four types of emotions(\emph{angry}, \emph{fear}, \emph{neutral}, \emph{sad}), but the recognition effect on \emph{happy} and \emph{surprise} is poor. Since  \emph{happy} is easily misclassified as \emph{sad}, it may be caused by the pendulum effect \cite{wegner1998putt} in psychology. Human emotions are characterized by multiplicity and bipolarity under the influence of external stimuli. Beside that, \emph{surprise} is always confused with \emph{fear}.  Due to the similar arousal \cite{hanjalic2005affective} of the two emotions, it may lead to them inducing each other.

\begin{table*}
\caption{Ablation experiments of different architectures for our model on IEMOCAP and CASIA, where '\cmark' or '\xmark' represents the network with or without the corresponding module. 'T-Trans', 'F-Trans', and 'TF-Trans' are the modules of Time Transformer, Frequency Transformer, and Time-frequency Transformer, respectively.}
\centering

\begin{tabular}{|c|c|c|c|c|c|c|c|}
\hline
\multirow{2}{*}{Architecture}& \multicolumn{3}{c|}{Ablation Experiments}&\multicolumn{2}{c|}{~~~~~IEMOCAP(\%)~~~~~}& \multicolumn{2}{c|}{~~~~~CASIA(\%)~~~~~} \\ \cline{2-8}
      & T-Trans & F-Trans & TF-Trans & ~~WAR~~      & ~~UAR~~       & ~~WAR~~      & ~~UAR~~            \\ \hline \hline
T$+$F  & \cmark    & \cmark    & \xmark    & 70.12        & 58.77        & 40.32        & 40.32               \\ \hline
T$+$TF  & \cmark   & \xmark    & \cmark    & 70.96        & 60.34         & 48.91        & 48.91                \\ \hline
F$+$TF  & \xmark     & \cmark    & \cmark    & 71.47        & 60.59         & 49.26        & 49.26               \\ \hline
T$+$F$+$TF (ours) & \cmark    & \cmark    & \cmark    &\textbf{74.43}&\textbf{62.90} &\textbf{53.17}&\textbf{53.17} \\ \hline
\end{tabular}
\label{tab:ablation-results}
\end{table*}

\subsubsection{Ablation Experiments}
We verified the effectiveness of our method by removing some modules of the proposed method. The experimental results are shown in Table~\ref{tab:ablation-results}, where 'T-Trans', 'F-Trans', and 'TF-Trans' are the modules of Time Transformer, Frequency Transformer, and Time-frequency Transformer, respectively. According to the results of the ablation experiments, the TF-Trans can effectively make the model focus on the emotion-related segments in the time-frequency domain and improve the emotional discrimination of the model. In addition, the effect of removing T-Trans is better than removing F-Trans, indicating that the frequency domain information of speech is of great significance for emotion recognition. Moreover, it is easy to observe that model\_1 achieves the worst result, particular on CASIA. Compared to model\_2 and model\_3, model\_1 has a significant performance degradation over 8\% of both WAR and UAR on CASIA. This phenomenon indicates the effectiveness of the Time-Frequency Transformer module. If we remove this part, the local time-domain and frequency-domain emotion features are not fully utilized to mine the time-frequency emotional correlations. Thus, the model cannot learn more discriminative global acoustic emotional feature representations.

\begin{figure*}[h]
\centering
\subfigure[Spectrogram]{
\label{Mel_spec_iemocap}
\includegraphics[width=0.4\linewidth]{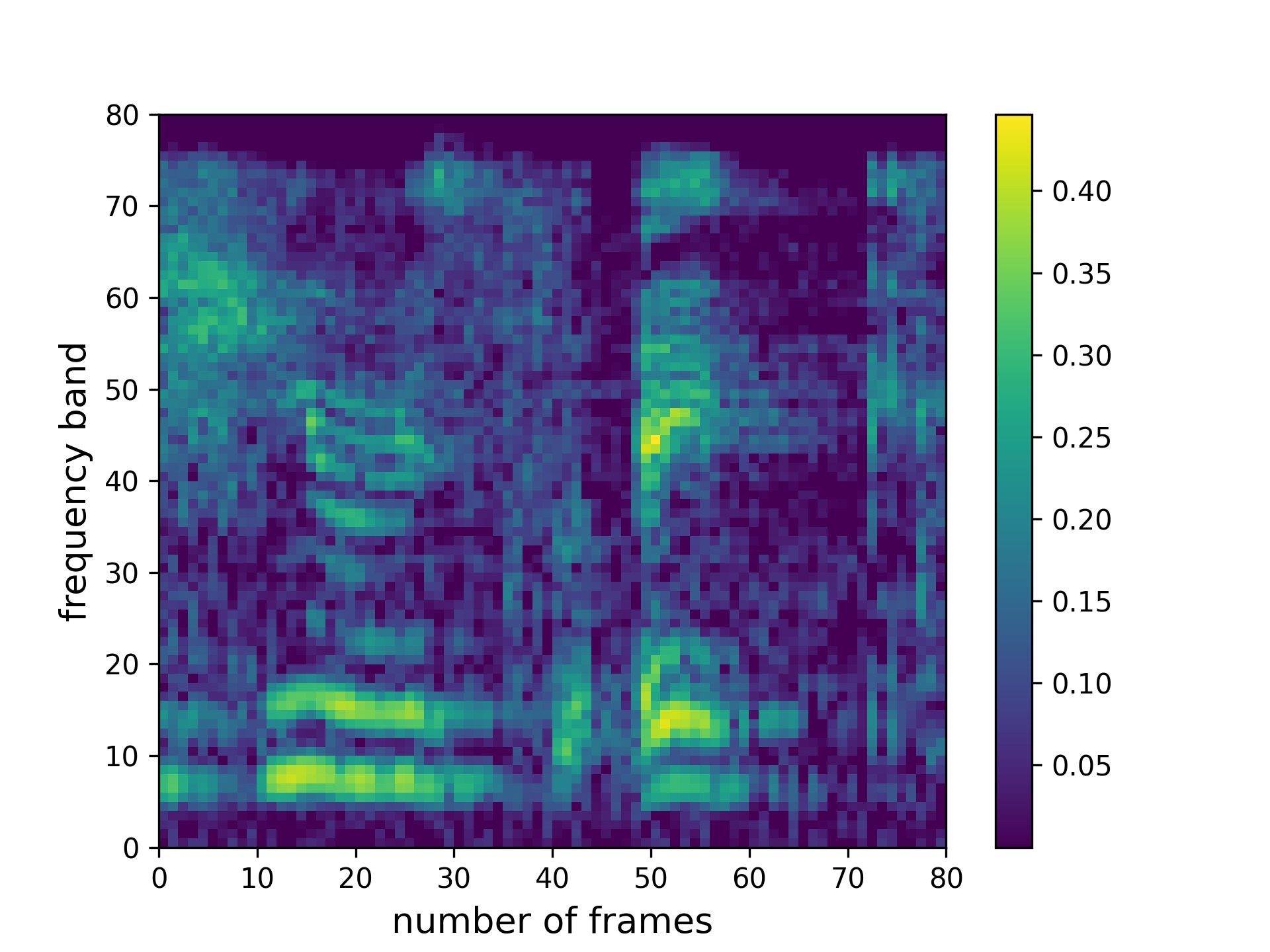}}
\subfigure[T-Trans Attention]{
\label{T_Trans_Attention_iemocap}
\includegraphics[width=0.4\linewidth]{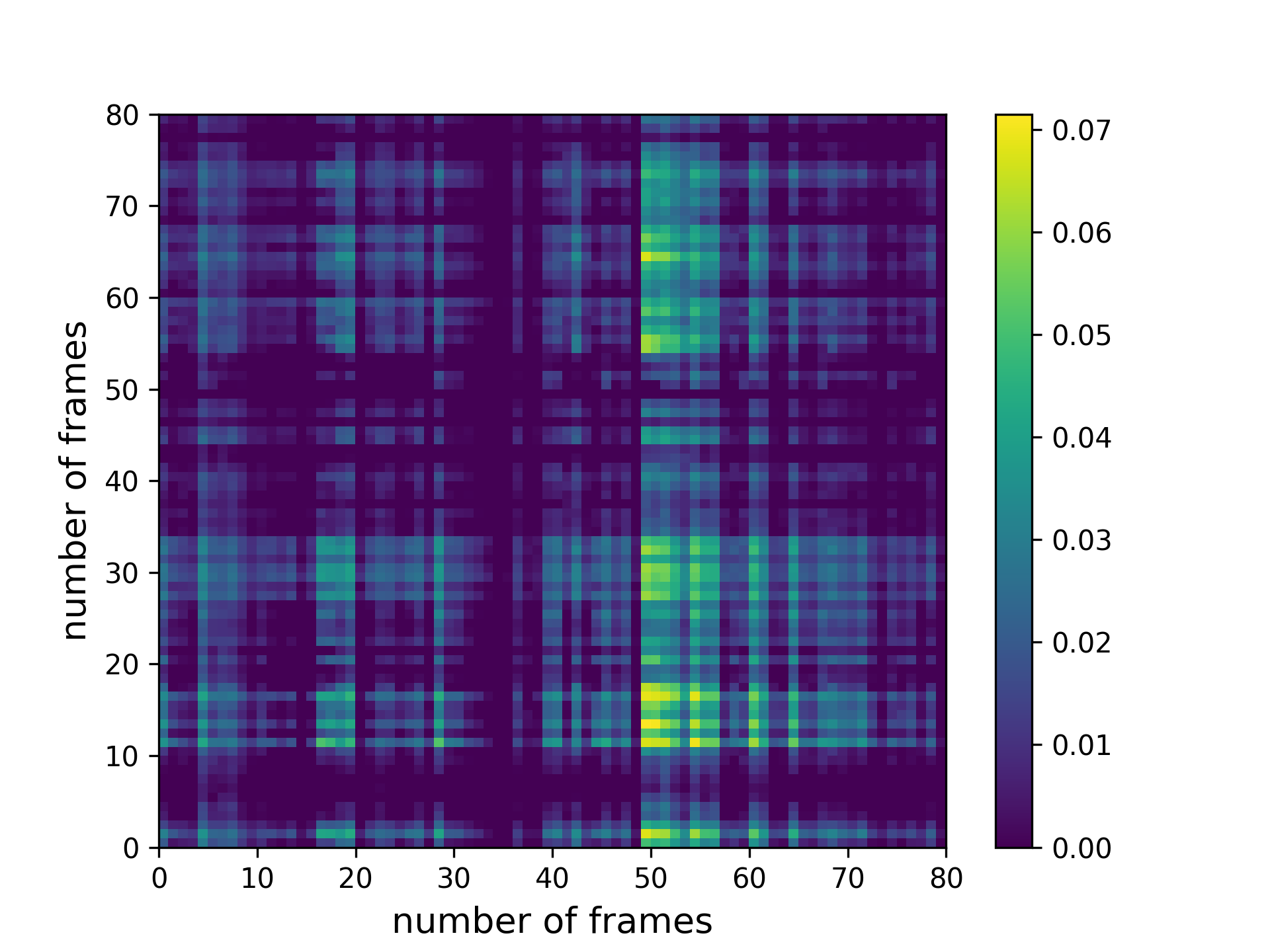}}
\subfigure[F-Trans Attention]{
\label{F_Trans_Attention_iemocap}
\includegraphics[width=0.4\linewidth]{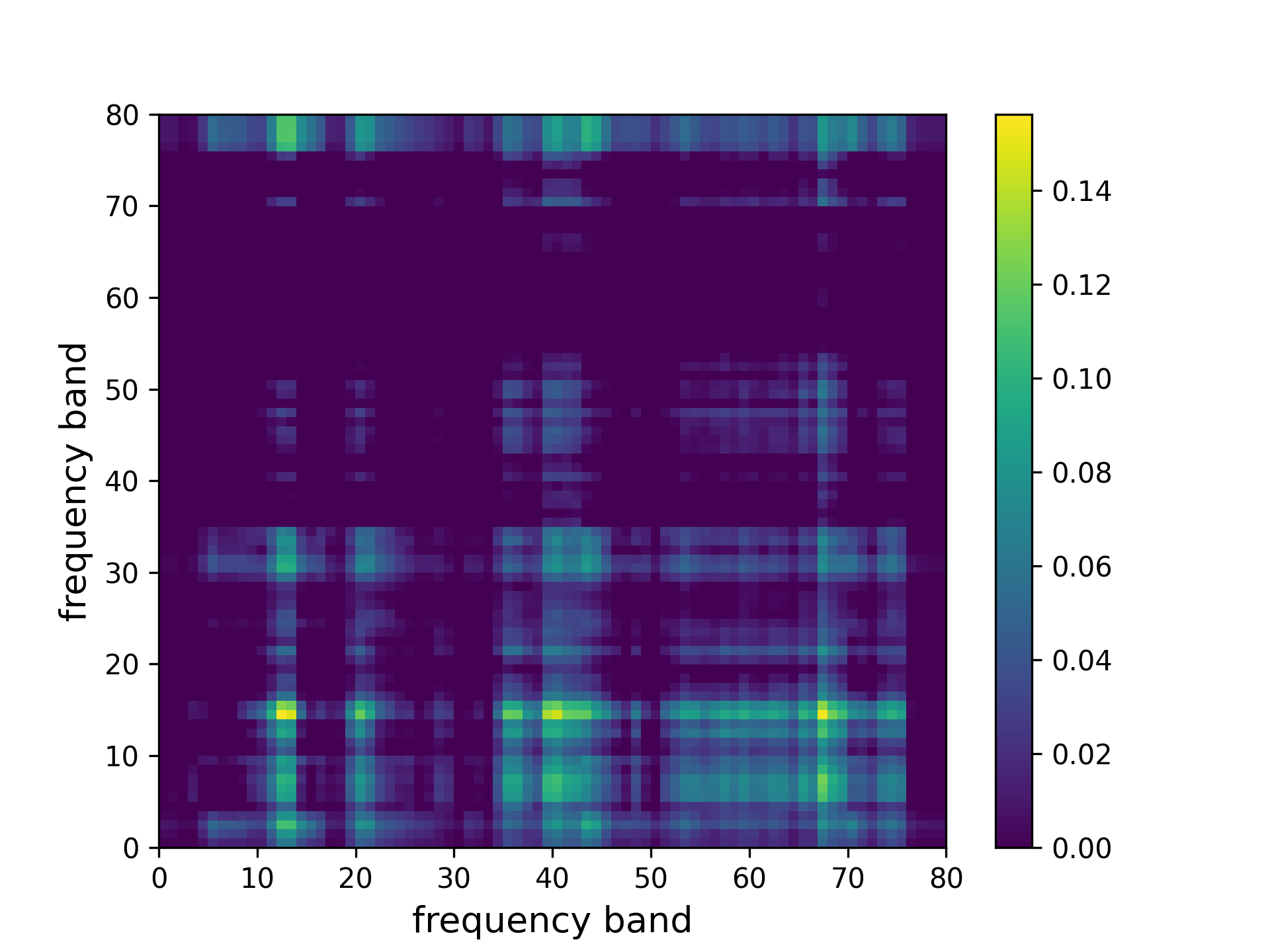}}
\subfigure[TF-Trans Attention]{
\label{TF_Trans_Attention_iemocap}
\includegraphics[width=0.4\linewidth]{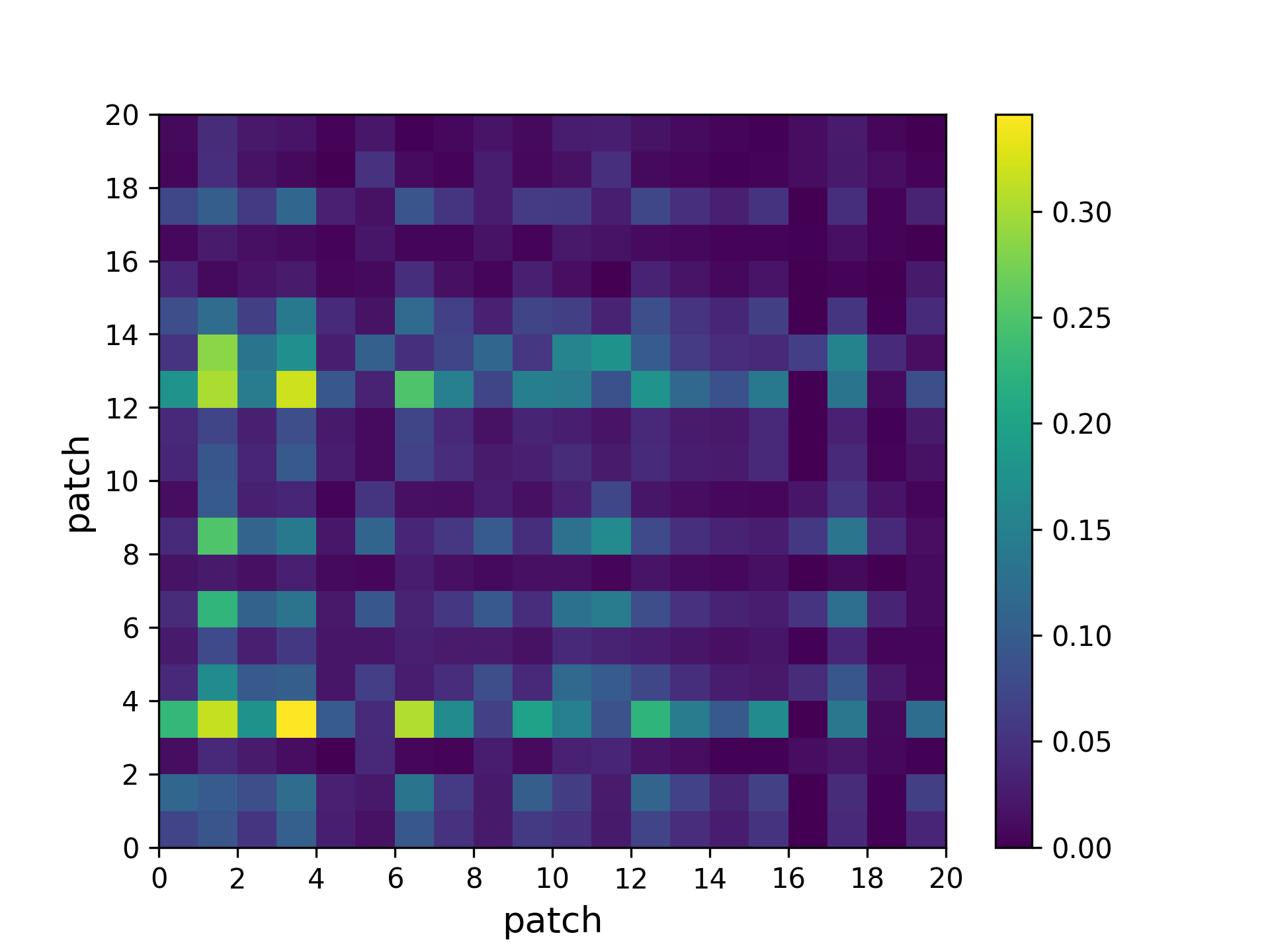}}
\caption{Visualization of log-Mel-spectrogram, T-Trans Attention, F-Trans Attention, and TF-Trans Attention (the emotion category of this sample is \emph{sad}, which belongs to IEMOCAP). }
\label{Visualization of Attention}
\end{figure*}


\subsubsection{Visualization of Attention}
In order to further investigate whether the proposed method focuses on frequency bands with specific energy activations and emotional key speech frames in the speech signal, we visualize the attention of Time Transformer, Frequency Transformer and Time-Frequency Transformer to the log-Mel-spectrogram, as shown in Figure. \ref{Visualization of Attention}. Form the visualization of Time Transformer in Figure. \ref{Visualization of Attention}(b), we can observe that there are strong activation values in $15^{th}-20^{th}$ frames and $50^{th}-60^{th}$ frames, which indicates that the emotion correlations between these frames is important to represent speech emotions. And these frames correspond to the positions with richer semantic information in Figure. \ref{Visualization of Attention}(a). The visualization of Frequency Transformer in Figure. \ref{Visualization of Attention}(c) shows that the activation of the middle and low frequency bands is more obvious, demonstrating that the middle and low frequency bands are key for the \emph{sad} emotion representation. From the results of time-frequency attention in Figure. \ref{Visualization of Attention}(d), we can see that the larger activation value (i.\,e., the salient patches) corresponds to the regions where the semantic information is more concentrated in Figure. \ref{Visualization of Attention}(a). Therefore, our proposed Time-Frequency Transformer can fully capture the time-frequency regions highly correlated with emotions while ensuring the complete modeling of local emotion information in the time and frequency domains to obtain discriminative speech emotion features.

\section{Conclusion}
In this paper, we propose a novel Transformer-based time frequency domain joint learning method for SER, i.e., Time-Frequency Transformer. It can effectively model local emotion correlations between frames and frequency bands through Time Frequency and Frequency Transformer. Then these local emotion features are aggregated into more discriminative global emotion representations by a Time-Frequency Transformer. However, the MSA operation in Transformer is aiming at model global long-range discrepancy, which is easily disturbed by noisy frames or frequency bands in speech. Therefore, our Future research will focus on sparse MSA for speech emotion representations.

%

%
%
%

\bibliographystyle{splncs04}
\bibliography{reference}

\end{document}